\documentclass[a4paper,11pt]{article}
\usepackage{pos}
\usepackage{aas_macros}

\title{HEGS : Revisiting a decade of H.E.S.S. extragalactic observations}

\author*[a]{François Brun}
\author[b]{David Sanchez}
\author[c]{Andrew M. Taylor}
\author[d]{Matteo Cerruti}
\author[e]{Jean-Philippe Lenain}
\onbehalf{on behalf of the H.E.S.S. collaboration}

\affiliation[a]{IRFU, CEA, Université Paris-Saclay,\\
                F-91191 Gif-sur-Yvette, France}

\affiliation[b]{Université Savoie Mont Blanc, CNRS, Laboratoire d’Annecy
                de Physique des Particules – IN2P3,\\ 
                74000 Annecy, France}
                
\affiliation[c]{Deutsches Elektronen-Synchrotron DESY,\\ 
                Platanenallee 6, 15738 Zeuthen, Germany}

\affiliation[d]{Université Paris Cité, CNRS, Astroparticule et Cosmologie,\\ 
                75013 Paris, France}

\affiliation[e]{Sorbonne Université, CNRS/IN2P3, Laboratoire de Physique Nucléaire et de Hautes Energies,\\ 
                LPNHE, 4 place Jussieu, 75005 Paris, France}

\emailAdd{francois.brun@cea.fr}

\abstract{During its first phase, from 2004 up to the end of 2012, the H.E.S.S. (High Energy Stereoscopic System) experiment observed the extragalactic skies for more than 2700 hours. These data have been re-analysed in a single consistent framework, leading to the derivation of a catalog of 23 sources. In total, about 5.7\% of the sky was observed, allowing for several additional studies to be conducted: source variability, extragalactic gamma-ray background light, and comparison with the \textit{Fermi}-LAT catalogues. In this contribution, we discuss these results and present the high-level data (catalogs, maps) released to the astrophysical community.}

\ConferenceLogo{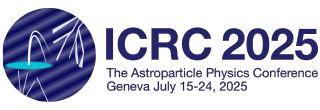}

\FullConference{39th International Cosmic Ray Conference (ICRC2025)\\
 15–24 July 2025\\
Geneva, Switzerland\\}

\begin{document}
\maketitle

\section{Introduction: The H.E.S.S. Extragalactic Sky Survey (HEGS)}

This contribution highlights the main results presented in \cite{hegs_paper} from the High Energy Stereoscopic System (H.E.S.S.) collaboration. H.E.S.S. is an array of Imaging Atmospheric Cherenkov Telescopes located in Namibia, observing the southern hemisphere sky in the very high-energy (VHE; $E > 100$ GeV) domain. This paper presents results covering the extragalactic observations of H.E.S.S. from 2004 to the end of 2012. 
In this time period, the array was in its initial configuration with four 12-meter diameter telescopes. 
The extragalactic observations strategy primarily involved pointed observations towards bright X-ray or \textit{Fermi}-LAT gamma-ray sources, alongside target-of-opportunity observations. This approach, while successful in detecting various extragalactic object classes, led to a highly non-uniform exposure distribution across the sky. The primary motivation for this work was to perform a consistent re-analysis of this large dataset.

\begin{figure*}
\centering
\includegraphics[width=0.95\textwidth]{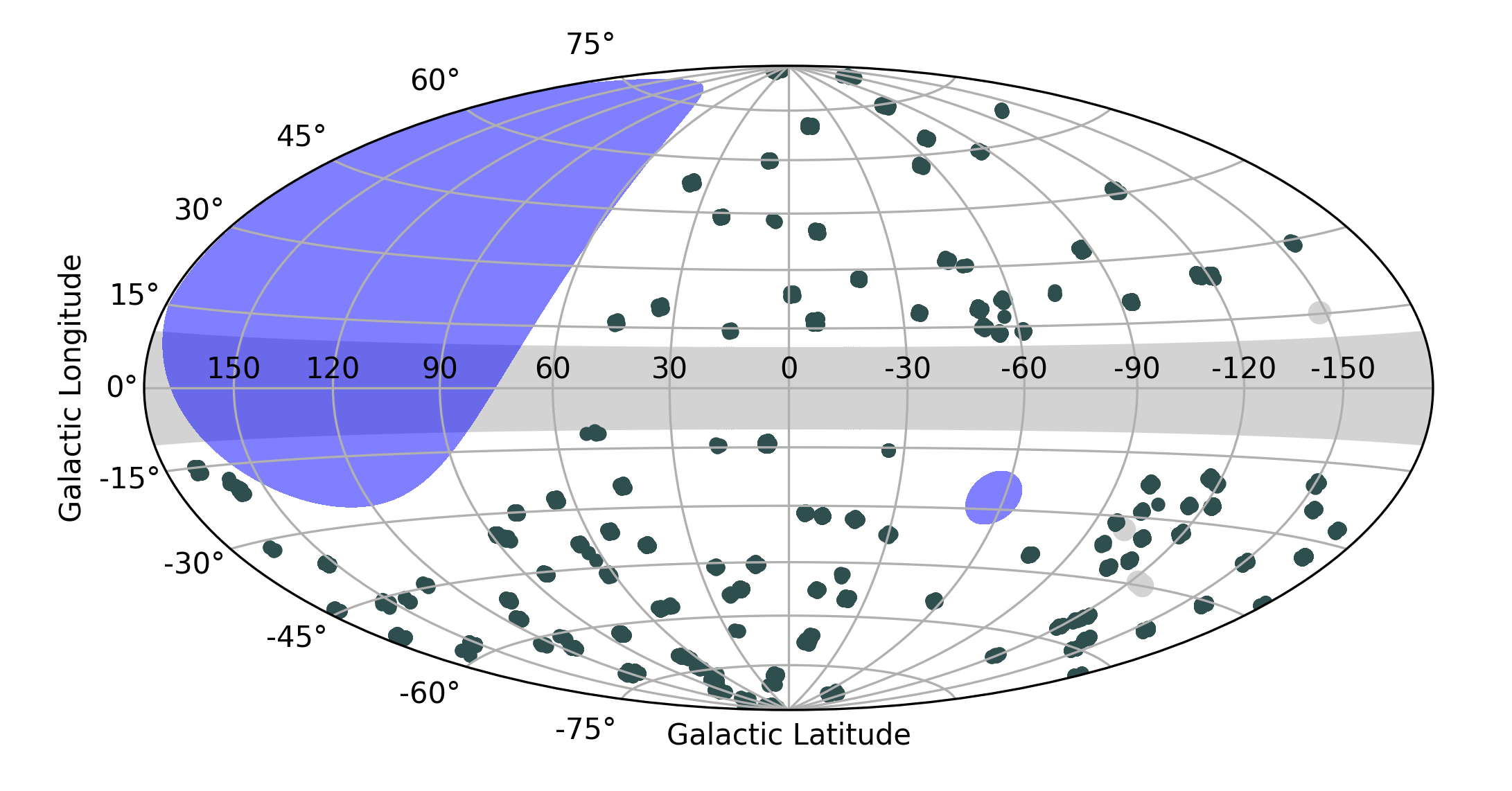}
\caption{Hammer-Aitoff sky map showing HEGS clusters (dark green). The grey band near the Galactic plane ($|b| < 10^{\circ}$) and other grey regions are excluded  from the analysis. Blue areas indicate sky regions unobservable by H.E.S.S. due to zenith angles above $60^{\circ}$. This figure is taken from \cite{hegs_paper}}
\label{fig_map}
\end{figure*}

Data selection ensured good atmospheric and instrumentation conditions, excluding observations at large zenith angles (>60°) to minimize systematic uncertainties. The data were grouped into 98 spatially separated observational clusters using the {\tt DBSCAN} algorithm \cite{10.1145/3068335} of the {\tt scikit-learn} library \cite{scikit-learn}, as presented on Fig. \ref{fig_map}. Regions within the Galactic plane ($|b| < 10^{\circ}$) and specific Galactic sources (e.g., Magellanic Clouds, 47 Tucanae) were discarded. These clusters collectively covered approximately 5.7\% of the total sky and accounted for 6500 observation runs, for a total of 2720 hours of observations.

\section{Analysis Methodology}

The data were analyzed using the likelihood reconstruction technique within the \textit{Model} framework \cite{Naurois}, optimized for a low energy threshold ("Loose cuts"). Sky maps were created with $0.01^{\circ}\times0.01^{\circ}$ pixels, where each pixel represented a test region for gamma-ray detection. The ring background method \cite{2007A&A...466.1219B} was employed to estimate the hadronic background, with known gamma-ray emitters masked to prevent contamination.

A source detection significance threshold of $5.7\sigma$ was determined using Monte Carlo simulations, ensuring a false detection probability of 3.89\% or less. For detected sources, a spectral analysis was conducted considering two models: a power-law (PL) and a log-parabola (LP). The preferred model was selected based on a likelihood ratio test (TS difference of 9, approximately $3\sigma$). Spectral results are affected by EBL absorption, which was corrected using the model from \cite{2011MNRAS.410.2556D}. A systematic uncertainty of 20\% on the flux and 0.2 on the spectral index are expected \cite{aha2006}.

In addition, variability searches were conducted on two timescales (run-wise and night-wise) using the ON-OFF test \cite{2020APh...11802429B}, fractional variability \cite{Vaughan}, and chi-squared fitting against a constant flux. Consistency checks with independent pipelines and archival H.E.S.S. publications confirmed the reliability of the results with differences generally within estimated systematic uncertainties.

\section{The HEGS Catalogue and Source Properties}

The HEGS analysis led to the detection of 23 sources, all of which were already known VHE gamma-ray emitters. The vast majority of these (18 out of 23) are BL Lacertae objects (BL Lacs). Other detected source types include two radiogalaxies (RGs; Centaurus~A and M~87, as well as PKS 0625-354 a debated RG/blazar), one flat-spectrum radio quasar (FSRQ; PKS 1510-089), and one starburst galaxy (NGC 253).

\begin{figure*}
\centering
        \includegraphics[width=0.99\textwidth]{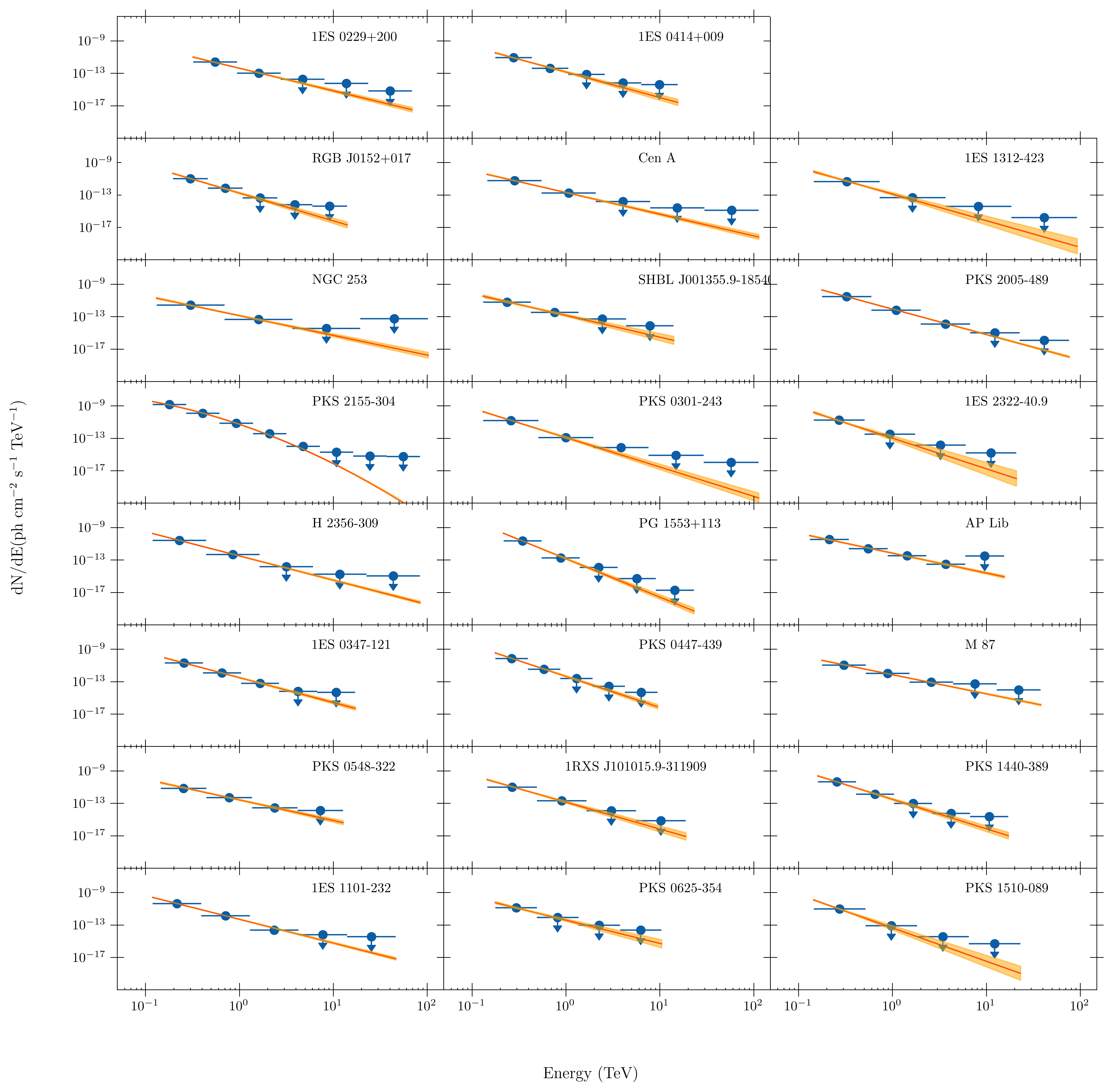}
    \caption{Spectra of detected HEGS sources. Red line: best-fit model; orange bands: 1$\sigma$ likelihood contours. No EBL correction is applied. The blue points indicate the flux points and the arrows indicate the 95\% CL upper limits computed for energy bins with a significance below $2\sigma$. This figure is taken from \cite{hegs_paper}}
    \label{fig_specs}
\end{figure*}

For spectral characterization, a power-law model was preferred for most sources, except for PKS 2155-304, which favored a log-parabola model. The spectrum obtained for each source is displayed in Fig. \ref{fig_specs}. 
A notable finding was the observation of a general spectral softening (or break) between the high-energy (HE, \textit{Fermi}-LAT) and VHE (H.E.S.S.) energy ranges, even when accounting for EBL absorption.

 Seven sources were identified as variable within the HEGS dataset. Most of these, such as M 87 \cite{2012ApJ...746..151A}, PG 1553+113 \cite{2015ApJ...802...65A}, PKS 2005-489 \cite{2011A&A...533A.110H}, and the highly variable PKS 2155-304 \cite{2017A&A...598A..39H}, were already known for their variability. However, RGB J0152+017 and 1ES 0347-121 showed newly detected variability due to the increased observation time in the HEGS dataset. H 2356-309 was also found to be variable on shorter (nightly) timescales than what was previously derived (monthly) \cite{2010A&A...516A..56H}.

 \section{Comparison with the \textit{Fermi}-LAT and Observational Bias}

By combining HE and VHE data, the spectral energy distributions (SEDs) of 15 non-variable sources are constrained using a log-parabola model. For most of these, the peak position in the SED was found to lie in the range of 10–100 GeV.

Comparisons with the \textit{Fermi}-LAT 4FGL catalogue \cite{2019arXiv190210045T} revealed that out of 247 HE emitters in the HEGS field-of-views, 12 sources were "constrained" by H.E.S.S. upper limits. This means that the extrapolated HE emission of these sources into the VHE domain predicted a flux higher than what H.E.S.S. observed, requiring the existence of a spectral break or cut-off in their spectra to reconcile the observations.

Simulations based on \textit{Fermi}-LAT luminosity functions demonstrated a significant observational bias of H.E.S.S. towards detecting BL Lac objects over FSRQs, with a simulated ratio of ~40 BL Lacs per FSRQ  at VHE. Given the instrument's energy threshold,  High-frequency peaked BL Lac (HBL) objects are preferentially detected. The observed distribution of spectral indices and redshifts for detected BL Lacs aligns well with these simulations.

The log N–log S distribution for BL Lac objects, which describes the cumulative number of sources above a given flux, yielded a best-fit index of $\gamma = 2.58 \pm 0.23$ for the number of sources per energy flux bin $dS$ approximated using the power-law relation $\frac{dN}{dS}\propto S^{-\gamma}$ \cite{2010ApJ...720..435A}. This value is consistent with the Euclidean expectation of 2.5, but also with values obtained when accounting for EBL attenuation and source evolution effects.

 \section{Contribution to the Extragalactic Gamma-Ray Background (EGB)} 

The study evaluated the contribution of blazars (BL Lacs and FSRQs) to the extragalactic gamma-ray background (EGB), integrating over their luminosity, redshift, and photon index distributions. It was found that BL Lac blazars dominate the high-energy end of the EGB (above ~1 GeV), while FSRQs dominate at lower energies. More specifically, HBLs are the dominant contributors to the EGB at energies above 100 GeV .

\begin{figure}
\centering
        \includegraphics[width=0.45\textwidth]{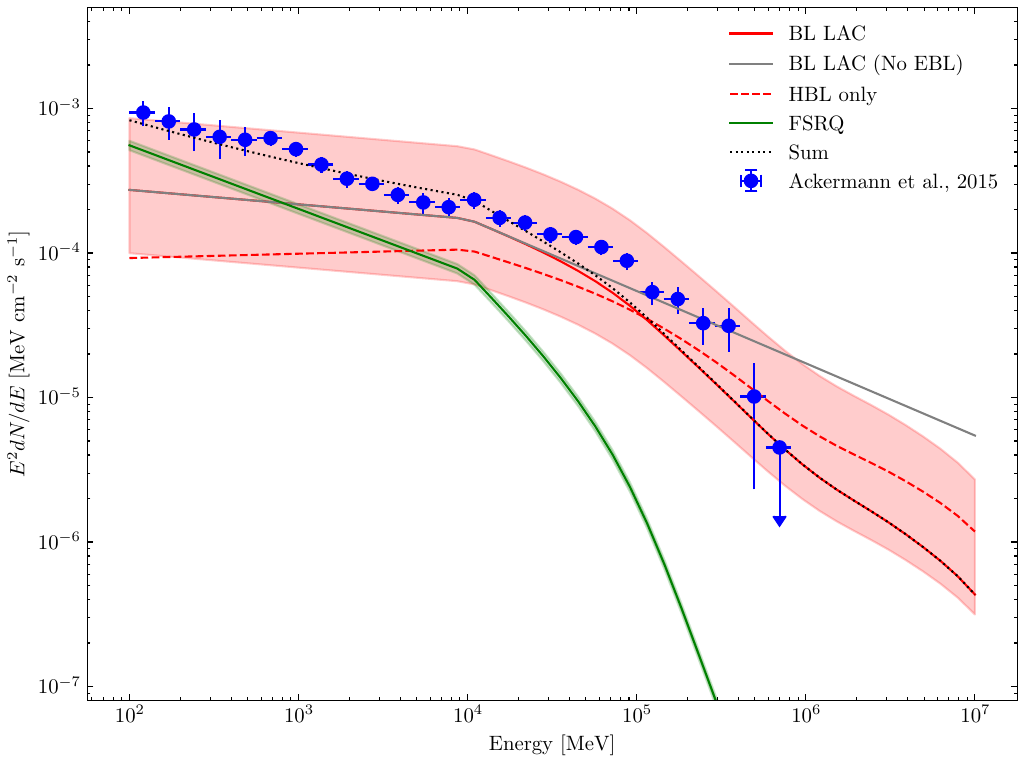}
        
    \caption{EGB measured by \textit{Fermi}-LAT (blue points). Red, grey, and green lines show contributions from BL Lac (with and without spectral break $\Delta \Gamma = 0.4$) and FSRQ, respectively. Shaded areas indicate uncertainties from luminosity function normalization. The dashed red line corresponds to the HBL luminosity function from \cite{2014ApJ...780...73A}. This figure is taken from \cite{hegs_paper}}
    \label{fig_egb}
\end{figure}

The current broken power-law model (assuming a break at $10$ GeV of $\Delta\Gamma=0.4$ for BL Lac blazars and $\Delta\Gamma=0.8$ for FSRQs), while consistent with \textit{Fermi}-LAT and H.E.S.S. observations, might lead to an over-contribution of HBLs to the EGB in the VHE band, potentially overshooting observed upper limits near 1 TeV, as visible in Fig. \ref{fig_egb}. Observations with H.E.S.S. phase II and the forthcoming Cherenkov Telescope Array Observatory (CTAO) \cite{2019scta.book.....C} are expected to provide more stringent constraints and refine the understanding of AGN contributions to the EGB, which is crucial for estimating the contribution from other messengers like ultra-high-energy cosmic ray (UHECR) cascades \cite{Taylor:2015rla} and neutrinos \cite{Fang:2022trf}.

 \section{Data Release}

As a significant outcome of this work, high-level data from the HEGS survey have been publicly released to the astrophysical community\footnote{\url{https://hess.science/pages/publications/auxiliary/2024_HEGS/}}. This includes FITS files containing descriptions of observations, the catalogue of detected sources, and several sky maps (significance, flux, upper limits, sensitivity).

\begin{figure*}
\centering
        \includegraphics[width=0.35\textwidth]{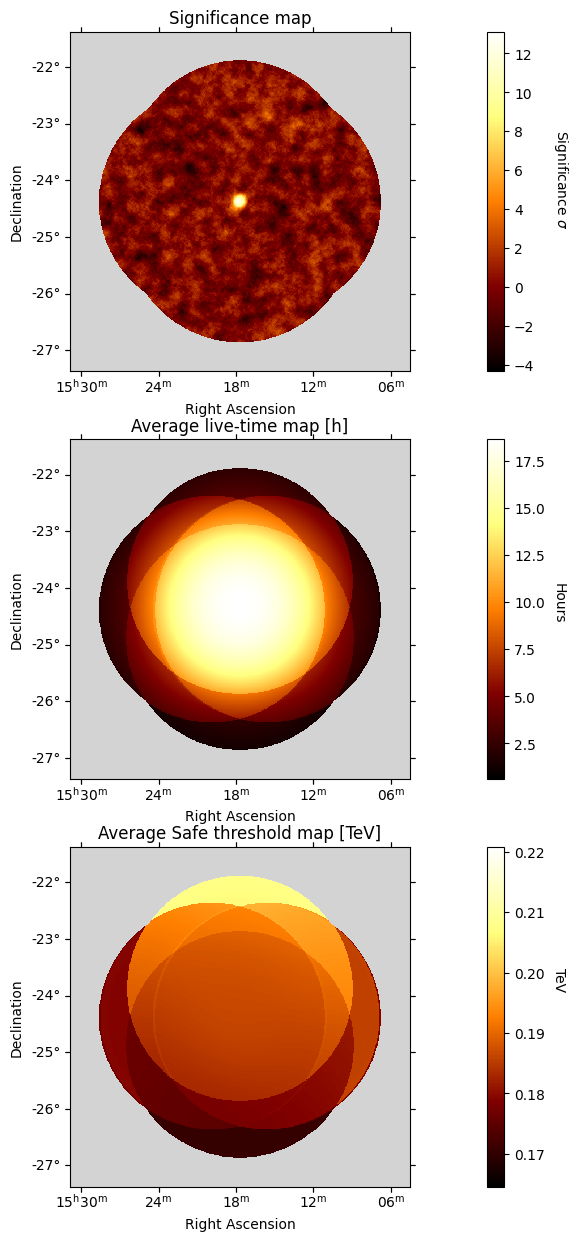}
        \includegraphics[width=0.45\textwidth]{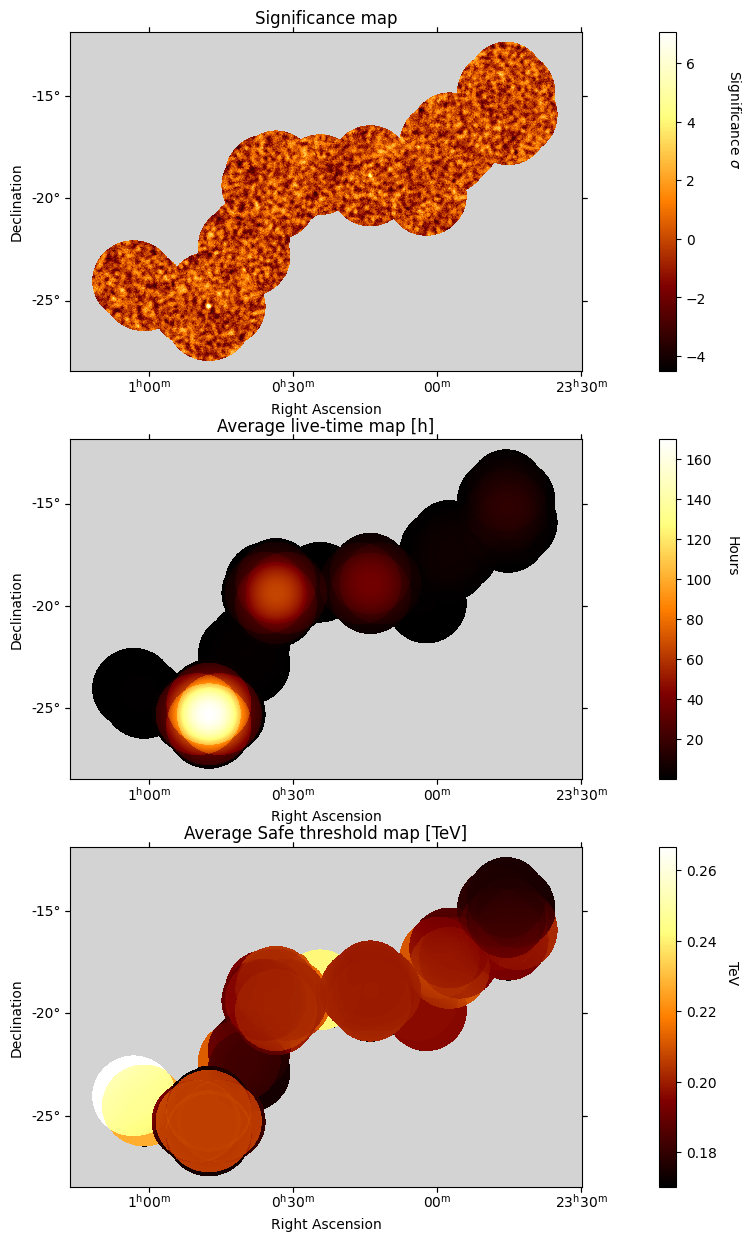}
    \caption{Example maps available from the catalogue FITS files: significance, average live time, and average energy threshold (top to bottom) for two observation clusters (left/right). This figure is taken from \cite{hegs_paper}.}
    \label{fig_maps}
\end{figure*}

As mentioned above, all the observations in this dataset have been grouped by clusters of observations. Each cluster is an independent sky region. The released sky maps were produced independently for each cluster, and each cluster is named according to the sky coordinates of its center. A table provides the position and spatial extent of each sky map. In addition, the observed live time per night for each cluster is also listed in a dedicated table. This information is indicative only: a cluster can be extended on the sky, therefore some observation time in a given night for a given cluster does not necessarily mean that the entire area of the cluster has been observed on that particular night. 

The HEGS catalogue contains information for each of the detected sources as well as the list of \textit{Fermi}-LAT sources within the HEGS FoV that are constrained by the H.E.S.S. observations. For the sources identified as variable, the lightcurve data (integral flux above $300$ GeV per night) are also provided.
For each cluster, a set of maps is released in a single FITS file : live time map, significance map, average energy threshold map, differential flux map, error map on the differential flux, flux upper limit map, and flux sensitivity maps at $5$ and $5.7\sigma$ levels.
Finally, a set of python scripts is provided to help with data exploration. These include examples demonstrating how to search for a sky position across the entire archive, retrieve relevant quantities if observation time is available, as well as how to display maps and source spectra from the catalog. Fig. \ref{fig_maps} shows examples of maps that are available from the archive.

\acknowledgments

This work was carried out within the framework of the H.E.S.S. Collaboration. Acknowledgments can be found on the following page: \url{https://hess.in2p3.fr/acknowledgements}.

\bibliographystyle{JHEP}
\bibliography{biblio_hegs_icrc25}

\end{document}